\documentclass[aps,prl,twocolumn,floats,epsfig,showpacs]{revtex4-1}
\usepackage{bbm}
\usepackage{amssymb}
\usepackage{ifpdf}
\usepackage{graphicx,epsfig}
\newcommand{\Z}{{\mathbb{Z}}}

\newcommand{\CP}{{\mathbb{C}P}}

\begin{document}

\title{Non-trivial $\theta$-Vacuum Effects in the 2-d $O(3)$ Model$^*$}

\author{M.\ B\"ogli$^a$, F.\ Niedermayer$^a$, M.\ Pepe$^b$, and 
U.-J.\ Wiese$^a$}

\affiliation{
$^a$ Albert Einstein Center for Fundamental Physics, Institute for Theoretical 
Physics, Bern University, Sidlerstr.\ 5, 3012 Bern, Switzerland \\
$^b$ INFN, Istituto Nazionale di Fisica Nucleare, Sezione di Milano-Bicocca
Edificio U2, Piazza della Scienza 3 - 20126 Milano, Italy \\
{$^*$ Dedicated to Peter Hasenfratz on the occasion of his 65th birthday}}

\begin{abstract}

We study $\theta$-vacua in the 2-d lattice $O(3)$ model using the standard 
action and an optimized constraint action with very small cut-off effects, 
combined with the geometric topological charge. Remarkably, dislocation lattice 
artifacts do not spoil the non-trivial continuum limit at $\theta \neq 0$, and 
there are different continuum theories for each value $0 \leq \theta \leq \pi$. 
A very precise Monte Carlo study of the step scaling function indirectly 
confirms the exact S-matrix of the 2-d $O(3)$ model at $\theta = \pi$.

\end{abstract} 


\maketitle

As Bethe showed in 1931, the spin $\frac{1}{2}$ antiferromagnetic Heisenberg
chain is gapless \cite{Bet31}. In 1983 Haldane conjectured that the spin $S$
chain has a gap for integer $S = 1,2,3,\dots$, but is gapless for half-integer 
spins $S = \frac{1}{2},\frac{3}{2},\frac{5}{2},\dots$ \cite{Hal83}. In the 
semi-classical large $S$ limit, he showed that the corresponding low-energy 
effective field theory is the 2-d $O(3)$ model at vacuum angle 
$\theta = 2 \pi S$, with coupling $g^2 = 2/S$. Chains with integer spin hence 
correspond to $\theta = 0$, while half-integer spin chains have $\theta = \pi$. 
The 2-d $O(3)$ model at $\theta = \pi$ should reduce to the $k=1$ WZNW model 
\cite{Wes71,Nov81,Wit84} at low energies. Haldane's conjecture has been 
confirmed by numerical simulations of integer and half-integer quantum spin 
chains \cite{Bot84,Sch95}. A direct comparison with the 2-d $O(3)$ model at 
$\theta = \pi$ is more difficult, due to the notorious complex action problem 
in numerical simulations. However, using a meron-cluster 
algorithm (an extension of the Wolff cluster algorithm \cite{Wol89}) it was 
possible to simulate at $\theta = \pi$, and consistency with the WZNW model 
predictions was obtained within statistical errors \cite{Bie95}. In this paper, 
we also use a variant of a method developed by Hasenbusch \cite{Has95,Bal10} to 
simulate $\theta$-vacuum effects in the 2-d $O(3)$ model with unprecedented per 
mille level precision. For the first time, this numerically confirms the 
conjectured exact S-matrix of the 2-d $O(3)$ model at $\theta = \pi$ 
\cite{Zam86} beyond any reasonable doubt, which also implies that the model 
indeed reduces to the WZNW model at low energies.

The 2-d $O(3)$ model is equivalent to the $\CP(1)$ model. 2-d $\CP(N-1)$ models
\cite{DAd78,Eic78} share many features with 4-d non-Abelian Yang-Mills theories:
they are asymptotically free, have a non-perturbatively generated massgap, 
instantons, as well as non-trivial $\theta$-vacua. $\CP(N-1)$ models on the 
lattice have been studied by Berg and L\"uscher \cite{Ber81} who introduced a 
geometric definition for the lattice topological charge $Q$. Field 
configurations with $Q = 1$ and a minimal value of the lattice action are known
as dislocations. In general, dislocations have a smaller action than $Q = 1$
instantons --- the minimal action configurations of the continuum theory. When
the dislocation action is less than a critical value (determined by the 1-loop
$\beta$-function coefficient), a semi-classical (but non-rigorous) argument 
suggests that the topological susceptibility $\chi_t = \langle Q^2 \rangle/V$ 
(where $V$ is the space-time volume) should suffer from an ultra-violet 
power-law 
divergence in the continuum limit \cite{Lue82}. In $\CP(N-1)$ models with 
$N \geq 4$, the dislocation problem does not arise when one uses the standard 
lattice action in combination with the geometric definition of the lattice 
topological charge. In the $\CP(2)$ model (i.e.\ for $N = 3$) the problem is 
expected to arise, but can be avoided by the use of an improved lattice action 
\cite{Lue83}, which pushes the dislocation action above the critical value. 
Finally, in the $\CP(1)$ (i.e.\ the $O(3)$) model the critical value of the 
dislocation action agrees with the continuum instanton action. Consequently, 
dislocations can be suppressed only by a delicate fine-tuning of the lattice 
action. This was realized in a study with a classically perfect lattice action, 
which indeed led to a logarithmic rather than a power-law divergence of 
$\chi_t$ \cite{Bla96}. Semiclassically, the logarithmic ultra-violet divergence 
also arises directly in the continuum \cite{Sch82,Lue82} and may thus not be a 
lattice artifact. In that case, $\chi_t$ would not be a physically meaningful 
quantity in the 2-d $O(3)$ model. Along the way, the whole concept of distinct 
topological sectors and corresponding $\theta$-vacua has been questioned in the 
2-d $O(3)$ model. In fact, one may suspect that $\theta$ is an irrelevant 
parameter that renormalizes to zero non-perturbatively. In this paper, for the 
first time we demonstrate with high accuracy, that $\theta$ is actually 
relevant and that each value $0 \leq \theta \leq \pi$ is associated with a 
different continuum theory.

A first indication that dislocations may not have a devastating effect on the 
continuum limit arose in a recent study of the 2-d $O(3)$ model at $\theta = 0$
with a topological lattice action \cite{Bie10}, where we found again just a
logarithmic divergence of $\chi_t$. Topological lattice actions are 
invariant against small deformations of the lattice fields. In particular, in
\cite{Bie10} an action with a constraint on the maximally allowed angle between 
neighboring spins has been used \cite{Pat92}. All allowed configurations (i.e.\ 
those that respect the constraint) have the same action value zero. As a 
consequence, this lattice model does not have the correct classical continuum 
limit, it violates the Schwarz inequality between action and topological charge,
and it cannot be treated in perturbation theory. Despite these various 
deficiencies, as was shown by very accurate Monte Carlo simulations of the step 
scaling function introduced in \cite{Lue91}, the 2-d $O(3)$ model with a 
topological lattice action still has the correct quantum continuum limit 
\cite{Bie10}. Furthermore, although there are even zero-action dislocations, 
$\chi_t$ was found to have only a logarithmic and not a power-law divergence. 
The continuum result for the step scaling function is known analytically thanks 
to an ingenious use of the thermodynamic Bethe ansatz \cite{Bal04}. Remarkably, 
the step scaling function is known analytically also at $\theta = \pi$ 
\cite{Bal11}. At intermediate values of $\theta$, on the other hand, the 2-d 
$O(3)$ model is expected not to be integrable. Still, by expanding around 
$\theta = \pi$, some interesting analytic results have been obtained even in 
that regime \cite{Con04}. 

As we will see, using very accurate Monte Carlo simulations, we confirm the 
analytic results for the step scaling function at $\theta = \pi$ with better 
than per mille level accuracy. This requires good control of lattice artifacts. 
In fact, in the 2-d $O(3)$ model (at $\theta = 0$) the behavior of lattice 
artifacts, which is apparently linear in the lattice spacing $a$, has been 
puzzling for many years \cite{Has02,Kne05}. Only recently, the puzzle has been
resolved by a careful analysis in the framework of Symanzik's improvement
program \cite{Bal09,Bal10}. It turned out that the apparent linear behavior is 
mimicked by the expected quadratic behavior modified by large logarithmic 
corrections. Interestingly, on some lattices the topological action approaches 
the continuum limit of the step scaling function from below, while the standard 
action approaches it from above. Here we combine both actions in such a way 
that cut-off effects are extremely reduced to at most a few per mille. Using 
that action as well as the standard action allows us to extrapolate reliably to 
the continuum limit.

Let us consider the $O(3)$ model on a triangulated square lattice as 
illustrated in Figure 1, with a 3-component unit-vector $\vec e_x \in S^2$
attached to each lattice site $x$. We choose periodic boundary conditions in 
the short spatial direction of extent $L$ and open boundary conditions in the 
long Euclidean time direction.
\begin{figure}[t]
\vskip-0.05cm
\includegraphics[width=0.5\textwidth]{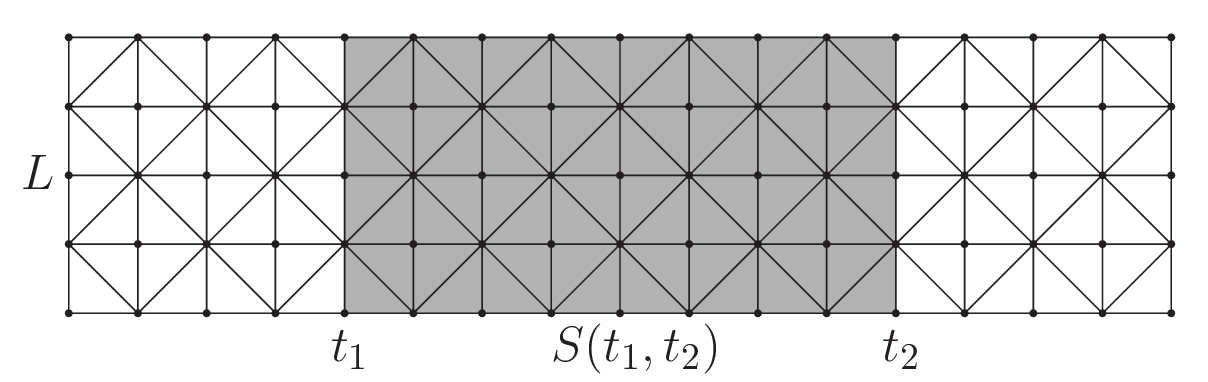}
\vskip-0.5cm
\caption{\it Triangulated square lattice: the triangles $\langle xyz \rangle$ 
in the shaded stripe $S(t_1,t_2)$ carry the topological term 
$i \theta q_{\langle xyz \rangle}$.}
\end{figure}
The action is defined on nearest-neighbor bonds $\langle xy \rangle$ (but not 
on the plaquette diagonals), and is given by 
\begin{equation}
S[\vec e] = \sum_{\langle xy \rangle} s(\vec e_x,\vec e_y), \quad
s(\vec e_x,\vec e_y) = \frac{1}{g^2} (1 - \vec e_x \cdot \vec e_y),
\end{equation} 
for $\vec e_x \cdot \vec e_y > \cos\delta$ and $s(\vec e_x,\vec e_y) = \infty$
otherwise. This action eliminates field configurations for which the angle
between neighboring spins exceeds $\delta$ \cite{Pat92}. For $\delta = \pi$ the 
constraint becomes irrelevant and the action reduces to the standard action, 
while for $g = \infty$ it reduces to the topological action of \cite{Bie10}. 
Besides the standard action, we will use an action with an optimized constraint 
angle, $\cos\delta = - 0.345$, which has very small cut-off effects 
\cite{Bal11a}.

Let us also define the geometric topological charge density 
$q_{\langle xyz \rangle} \in [- \frac{1}{2},\frac{1}{2}]$ associated with a 
triangle $\langle xyz \rangle$,
\begin{eqnarray}
R \exp(2 \pi i q_{\langle xyz \rangle})&=&1 + \vec e_x \cdot \vec e_y +
\vec e_y \cdot \vec e_z + \vec e_z \cdot \vec e_x \nonumber \\
&+&i \vec e_x \cdot (\vec e_y \times \vec e_z), \quad R \geq 0,
\end{eqnarray}
with $4 \pi q_{\langle xyz \rangle}$ being the oriented area of the spherical 
triangle on $S^2$ defined by the three unit-vectors $\vec e_x$, $\vec e_y$, and 
$\vec e_z$. On a periodic lattice, the topological charge 
$Q = \sum_{\langle xyz \rangle} q_{\langle xyz \rangle}$
would be an integer in the second homotopy group $\Pi_2[S^2] = \Z$. Here we work
with open boundary conditions in Euclidean time and we sum the topological 
charge density only over the stripe $S(t_1,t_2)$ of shaded plaquettes between 
$t_1$ and $t_2$ illustrated in Figure 1. This yields the non-integer valued 
quantity
$Q(t_1,t_2) = \sum_{\langle xyz \rangle \in S(t_1,t_2)} q_{\langle xyz \rangle}$. In order 
to determine the massgap, we introduce the operator 
$\vec E(t) = \sum_{x_1} \vec e_x$, where the sum extends over all points 
$x = (x_1,t)$ in a time-slice. We now define the 2-point function
\begin{eqnarray}
&&C(t_1,t_2;\theta) = \frac{1}{Z(t_1,t_2;\theta)} \prod_x \int_{S^2} d\vec e_x \ 
\vec E(t_1) \cdot \vec E(t_2) \nonumber \\
&&\times \exp(- S[\vec e] + i \theta Q(t_1,t_2)) \sim
\exp(- m(\theta,L)(t_2-t_1)), 
\nonumber \\
&&Z(t_1,t_2;\theta) = \prod_x \int_{S^2} d\vec e_x \ 
\exp(- S[\vec e] + i \theta Q(t_1,t_2)),
\end{eqnarray}
which decays exponentially with the $\theta$- and $L$-dependent 
massgap $m(\theta,L)$ at large Euclidean time separations. The massgap has been 
determined with high accuracy from numerical simulations using the meron-cluster
algorithm \cite{Bie95} as well as a variant of a method developed by Hasenbusch 
\cite{Has95}, which was further improved in \cite{Bal09}. It is straightforward 
to include the topological term in this method. Numerically, we obtain 
$C(t_1,t_2;\theta)$ as the ratio of $C(t_1,t_2;\theta) Z(t_1,t_2;\theta)/Z(0)$ 
and $Z(t_1,t_2;\theta)/Z(0)$, where $Z(0) = Z(t_1,t_2;0)$, which is independent 
of $t_1$ and $t_2$. It turns out that the resulting complex action problem for 
$\theta \neq 0$ is mild for the moderate spatial volumes that are relevant in 
this study.

Following \cite{Lue91}, we set $u_0 = L m(\theta,L)$ and define the
step 2 scaling function $\Sigma(\theta,2,u_0,a/L) = 2 L m(\theta,2 L)$. 
Remarkably, both for $\theta = 0$ \cite{Bal04} and for $\theta = \pi$ 
\cite{Bal11}, the continuum limit 
$\sigma(\theta,2,u_0) = \Sigma(\theta,2,u_0,a/L \rightarrow 0)$ has been 
determined analytically. One particular value is 
$\sigma(\theta = 0,2,u_0 = 1.0595) = 1.2612103$ \cite{Bal04}. Figure 2
shows the cut-off effects of $\Sigma(0,2,u_0,a/L)$ for the standard action, the
topological action, and the optimized constraint action. The constraint 
has been optimized to $\cos\delta = - 0.345$ by demanding that 
$\Sigma(0,2,u_0,a/L=\frac{1}{10}) = \sigma(0,2,u_0)$ for $u_0 = 1.0595$. For 
$L/a \geq 6$, the remaining cut-off effects are then less than a per mille. The 
data are fitted to Symanzik's effective theory, which predicts \cite{Bal09}
\begin{eqnarray}
\label{Sym}
&&\Sigma(0,2,u_0,a/L) = \nonumber \\
&&\sigma(0,2,u_0) + 
\frac{a^2}{L^2} \left[B \log^3(L/a) + C \log^2(L/a) + \dots\right].
\end{eqnarray}
They are in excellent agreement with the analytic prediction to four significant
digits accuracy.
\begin{figure}[t]
\begin{center}
\includegraphics[width=0.47\textwidth,angle=0]{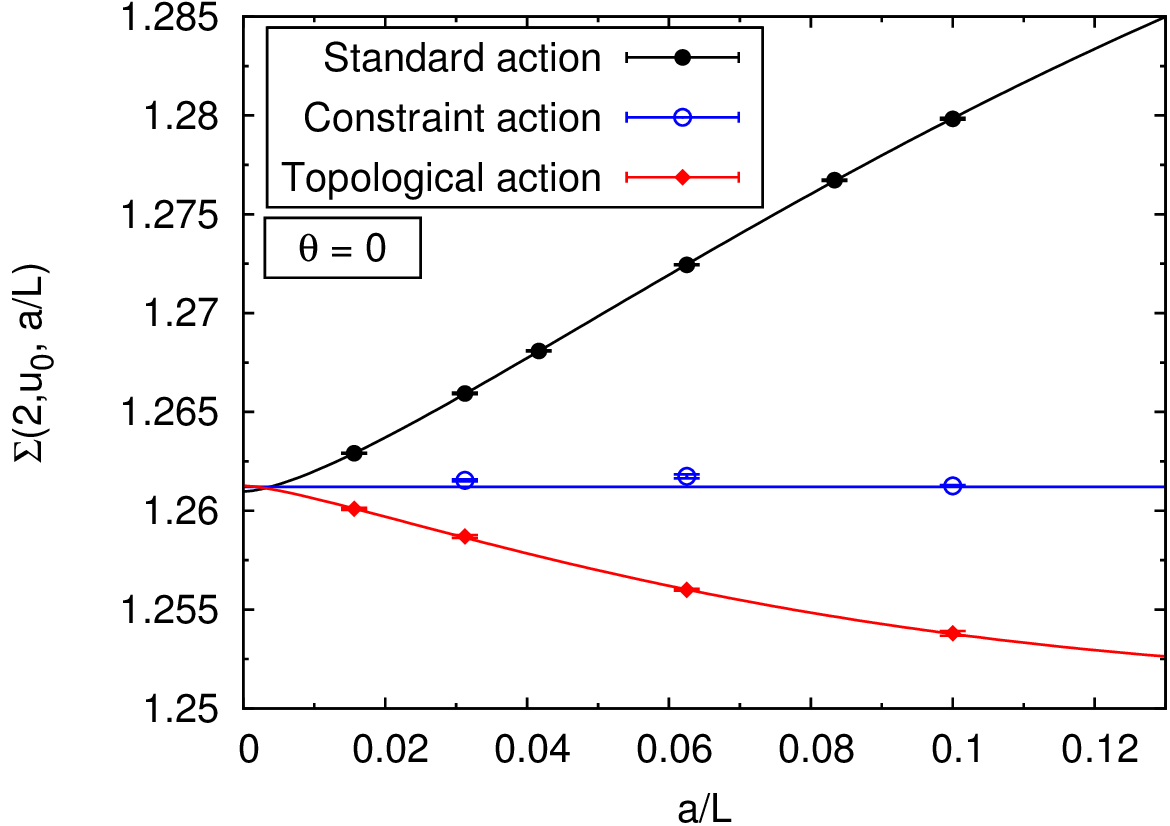} \\
\includegraphics[width=0.47\textwidth,angle=0]{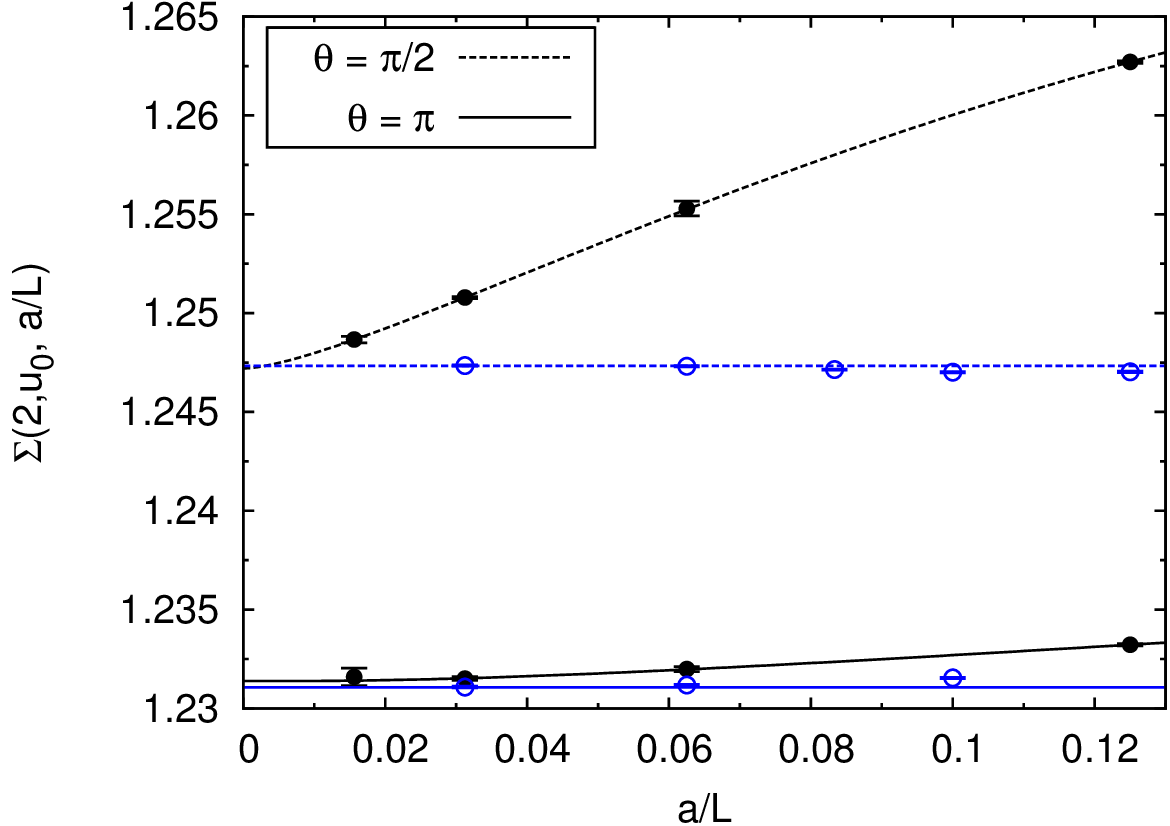}
\caption{\it Cut-off dependence of the step scaling function 
$\Sigma(\theta,2,u_0 = 1.0595,a/L)$ for three different lattice 
actions: the standard action, the topological lattice action of \cite{Bie10},
and the optimized constraint action with $\cos\delta = - 0.345$, as well as for 
three different values of $\theta = 0$ (top) and $\theta = \frac{\pi}{2}, \pi$
(bottom). The lines are 
fits based on eq.(\ref{Sym}). The horizontal lines represent the analytic 
continuum results for $\theta = 0$ \cite{Bal04} and $\theta = \pi$ 
\cite{Bal11}, and the fitted continuum value for $\theta = \frac{\pi}{2}$.}
\end{center}
\end{figure}

Figure 2 also shows corresponding results at $\theta = \pi$ and $\pi/2$, both 
for the standard action and for the optimized constraint action with 
$\cos\delta = - 0.345$. At $\theta = \pi$ the analytic result for the step
scaling function is $\sigma(\pi,2,u_0 = 1.0595) = 1.231064$ \cite{Bal11}.
Remarkably, although the constraint was optimized for $\theta = 0$, the cut-off
effects are still at most a few per mille at $\theta = \pi$. Since 
$\exp(i \pi Q) = (-1)^Q = \exp(- i \pi Q)$, for $\theta = \pi$ the topological 
term does not explicitly break any symmetries of the $\theta = 0$ theory.
Consequently, the corresponding analysis of the cut-off effects \cite{Bal09},
which underlies the fits in Figure 2, still applies. Again, we find excellent
agreement with the analytic result to four digits accuracy, which 
indirectly confirms the corresponding exact S-matrix \cite{Zam86} that 
underlies the analytic calculation. Agreement of the same quality is obtained 
for other values of $u_0$.

Finally, let us consider the case $\theta = \frac{\pi}{2}$, which is again
illustrated in figure 2. For $\theta \neq 0$ or $\pi$ there are no analytic 
results. The topological term then breaks both parity and charge conjugation 
and hence the detailed analysis of the cut-off effects in \cite{Bal09} no 
longer applies. Still, remarkably the optimized constraint action with 
$\cos\delta = - 0.345$ has no observable cut-off effects, and extrapolates to 
the same continuum value $\sigma(\frac{\pi}{2},2,u_0 = 1.0595) = 1.24733(4)$ 
as the standard action. This value of the step scaling function differs from
the values at $\theta = 0$ and $\theta = \pi$ by three significant digits. This
shows that the continuum limit at $\theta = \frac{\pi}{2}$ indeed represents a 
different theory. The same is true for other intermediate values 
$0 < \theta < \pi$. Hence, $\theta$ is indeed a relevant physical parameter 
that does not renormalize to 0 or $\pi$ non-perturbatively, as one might have
expected due to the presence of dislocations.

If $\chi_t$ is logarithmically ultra-violet divergent in the continuum limit,
which may not be the case for other definitions of the topological charge
\cite{Lue04}, the difference between the energy densities of different 
$\theta$-vacua diverges as well. This is a peculiarity of the 2-d $O(3)$ model, 
which should not extend to $\CP(N-1)$ models with $N>2$. Despite the divergence 
of the vacuum energy density, the massgap $m(\theta,L)$ is completely
well-behaved, and a proper non-trivial continuum field theory can be defined 
for any value of $\theta$. It will be interesting to further investigate these 
theories. At large volume $L$ and low energies, for $\theta = \pi$ one expects 
the $O(3)$ symmetry to dynamically enhance to the 
$O(4) = SU(2)_L \times SU(2)_R$ symmetry of the WZNW model. In the infinite 
volume limit, the triplet state should then become massless, i.e.\ 
$m(\theta = \pi,L \rightarrow \infty) \rightarrow 0$, and degenerate with an 
$O(3)$ singlet state. For small values of $\theta$, the singlet state is a 
scattering state with an energy $2 m(\theta,L \rightarrow \infty)$. For 
example, it would be interesting to figure out whether there is a critical 
value $\theta_c$ at which the singlet becomes a bound state \cite{Con04}. Since
$\theta$ does not get renormalized, despite the fact that the energy density
difference between different $\theta$-vacua may be logarithmically ultra-violet
divergent, the massgap is a finite physical quantity, whose value at 
$\theta = \pi$ is even known analytically; e.g.\ at $L m(0,L) = 1.0595$ one 
obtains $L m(\pi,L) = 1.048175$ \cite{Bal11}. As shown in Figure 3, the 
numerical data for $L = 24 a$, which extend to arbitrary values of $\theta$, 
agree with this analytic prediction below the permille level. There is
a remaining tiny cut-off effect, which diminishes only for volumes as large as
$L = 128 a$. The cut-off effects of constraint actions will be discussed in
detail in forthcoming publications \cite{Bal11a}. 
\begin{figure}[t]
\begin{center}
\includegraphics[width=0.48\textwidth]{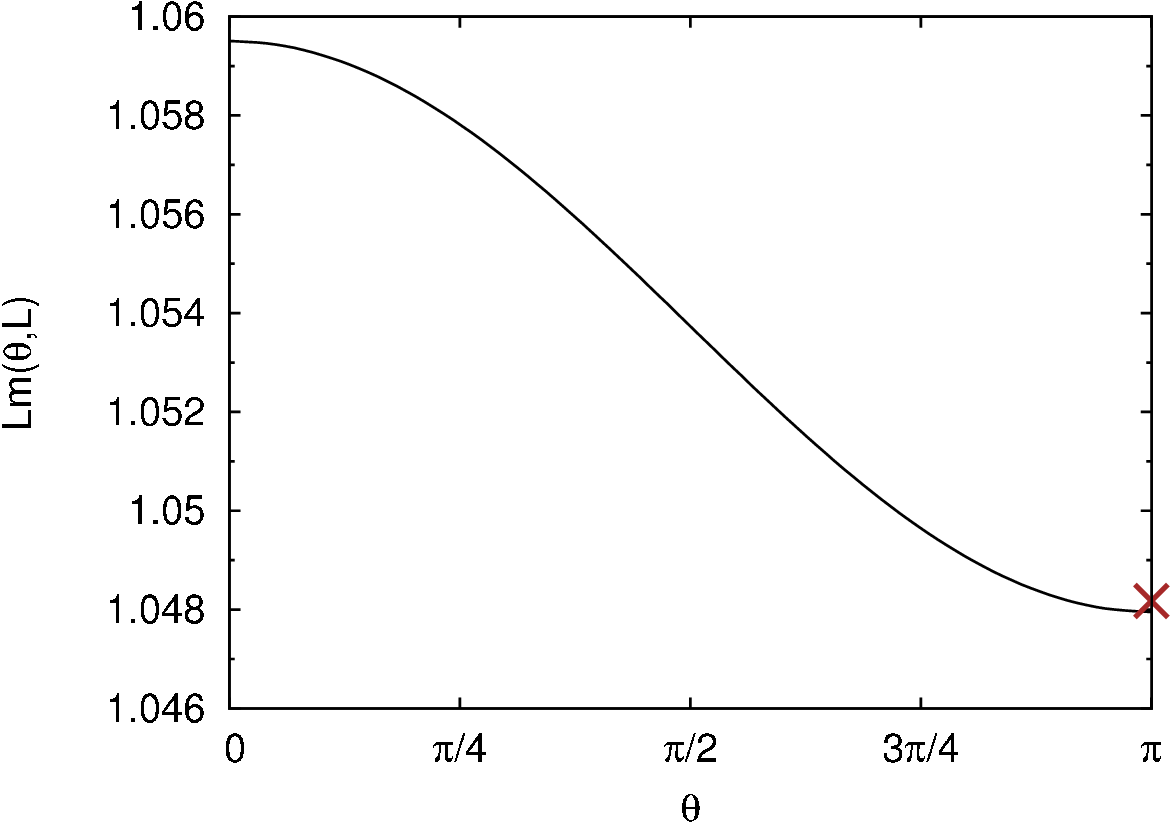}
\caption{\it The $\theta$-dependent massgap $L m(\theta,L)$ at 
$L m(0,L) = 1.0595$ using the optimized constraint action for $L = 24 a$, 
compared to the analytic result at $\theta = \pi$ \cite{Bal11} (cross).}
\end{center}
\end{figure}

Since dislocations do not prevent a non-trivial continuum limit in the 2-d 
$O(3)$ model with $\theta \neq 0$, the same is expected for 2-d $\CP(N-1)$ 
models. Using an unconventional D-theory regularization by $SU(N)$ 
quantum spin ladders, $\CP(N-1)$ models have been simulated successfully at 
$\theta = \pi$ \cite{Bea05}. More recently, a worm algorithm has been 
constructed for $\CP(N-1)$ models \cite{Wol10}, which, however, suffers from a 
sign problem at $\theta \neq 0$. Still, this algorithm or a variant of 
Hasenbusch's method may be sufficiently powerful to simulate $\CP(N-1)$ models 
for arbitrary values of $\theta$, and thus generalize the results we have 
obtained here. Furthermore, it is obvious to ask whether $\theta$-vacua effects 
in 4-d Yang-Mills theories can be addressed with similar methods, using a 
geometric definition of the lattice topological charge 
\cite{Lue82a,Phi86,Goe86,Goe87}. In that case, a potential dislocation problem 
arises for $SU(2)$ and $SU(3)$ \cite{Pug89}, but can be cured by using an 
improved lattice action \cite{Goe89}. Based on the results obtained here, we 
expect that dislocations are harmless also in 4-d Yang-Mills theories, and that
$\theta$-vacuum effects can be simulated at least in moderate spatial volumes.

We dedicate this paper to P.\ Hasenfratz on the occasion of his 65th birthday.
Over many years, we have benefitted tremendously from his deep insights into
strongly interacting field theories, which he generously shared with us.
We are indebted to J.\ Balog, M.\ L\"uscher, and P.\ Weisz for illuminating 
discussions, and to J.\ Balog for providing exact values of the step scaling 
function at $\theta = \pi$ prior to publication. This work has been supported 
by the Regione Lombardia and CILEA Consortium through a LISA 2011 grant, as 
well as by the Schweizerischer Nationalfonds (SNF).


\begin{thebibliography}{20}

\bibitem{Bet31}
H.\ Bethe, Z.\ Phys.\ A71 (1931) 205.

\bibitem{Hal83}
F.\ D.\ M.\ Haldane, Phys.\ Rev.\ Lett.\ 50 (1983) 1153.

\bibitem{Wes71}
J.\ Wess and B.\ Zumino, Phys.\ Lett.\ B37 (1971) 95.

\bibitem{Nov81}
S.\ P.\ Novikov, Sov.\ Math.\ Dokl.\ 24 (1981).

\bibitem{Wit84}
E.\ Witten, Commun.\ Math.\ Phys.\ 92 (1984) 455.

\bibitem{Bot84}
R.\ Botet, R.\ Jullien, and M.\ Kolb, Phys.\ Rev.\ B30 (1984) 215.

\bibitem{Sch95}
U.\ Schollw\"ock and T.\ Jolicoeur, Europhys.\ Lett.\ 30 (1995) 493.

\bibitem{Wol89}
U.\ Wolff, Phys.\ Rev.\ Lett.\ 62 (1989) 361.

\bibitem{Bie95}
W.\ Bietenholz, A.\ Pochinsky, and U.-J.\ Wiese, Phys.\ Rev.\ Lett.\ 75 (1995) 
4524.

\bibitem{Has95}
M.\ Hasenbusch, Nucl.\ Phys.\ Proc.\ Suppl.\ 42 (1995) 764.

\bibitem{Bal10}
J.\ Balog, F.\ Niedermayer, and P.\ Weisz, Nucl.\ Phys.\ B824 (2010) 563.

\bibitem{Zam86}
A.\ B.\ Zamolodchikov and V.\ A.\ Fateev, Sov.\ Phys.\ JETP 63 (1986) 913.

\bibitem{DAd78}
A.\ D'Adda, P.\ Di Vecchia, and M.\ L\"uscher, Nucl.\ Phys.\ B146 (1978) 63;
Nucl.\ Phys.\ B152 (1979) 125.

\bibitem{Eic78}
H.\ Eichenherr, Nucl.\ Phys.\ B146 (1978) 215.

\bibitem{Ber81}
B.\ Berg and M.\ L\"uscher, Nucl.\ Phys.\ B190 (1981) 412.

\bibitem{Lue82}
M.\ L\"uscher, Nucl.\ Phys.\ B200 (1982) 61.

\bibitem{Lue83}
M.\ L\"uscher and D.\ Petcher, Nucl.\ Phys.\ B225 (1983) 53.

\bibitem{Bla96}
M.\ Blatter, R.\ Burkhalter, P.\ Hasenfratz, and F.\ Niedermayer, 
Phys.\ Rev.\ D53 (1996) 923.

\bibitem{Sch82}
P.\ Schwab, Phys.\ Lett.\ B118 (1982) 373.

\bibitem{Bie10}
W.\ Bietenholz, U.\ Gerber, M.\ Pepe, and U.-J.\ Wiese, JHEP 1012 (2010) 020.
4524.

\bibitem{Pat92}
A.\ Patrascioiu and E.\ Seiler, J.\ Stat.\ Phys.\ 69 (1992) 573;
Nucl.\ Phys.\ Proc.\ Suppl.\ 30 (1993) 184. 

\bibitem{Lue91}
M.\ L\"uscher, P.\ Weisz, and U.\ Wolff, Nucl.\ Phys.\ B359 (1991) 221.

\bibitem{Bal04}
J.\ Balog and A.\ Hegedus, J.\ Phys. A37 (2004) 1881; 
Nucl.\ Phys.\ B725 (2005) 531; Nucl.\ Phys.\ B829 (2010) 425.

\bibitem{Bal11}
J.\ Balog, private communication.

\bibitem{Con04}
D.\ Controzzi and G.\ Mussardo, Phys.\ Rev.\ Lett.\ 92 (2004) 021601.

\bibitem{Has02}
M.\ Hasenbusch, P.\ Hasenfratz, F.\ Niedermayer, B.\ Seefeld, and U.\ Wolff,
Nucl.\ Phys.\ Proc.\ Suppl.\ 106 (2002) 911.

\bibitem{Kne05}
F.\ Knechtli, B.\ Leder, and U.\ Wolff, Nucl.\ Phys.\ B726 (2005) 421.

\bibitem{Bal09}
J.\ Balog, F.\ Niedermayer, and P.\ Weisz, Phys.\ Lett.\ B676 (2009) 188.

\bibitem{Bal11a}
J.\ Balog, F.\ Niedermayer, M.\ Pepe, P.\ Weisz, and U.-J.\ Wiese, in 
preparation.

\bibitem{Lue04}
M.\ L\"uscher, Phys.\ Lett.\ B593 (2004) 296.

\bibitem{Bea05}
B.\ B.\ Beard, M.\ Pepe, S.\ Riederer, and U.-J.\ Wiese, Phys.\ Rev.\ Lett.\ 94
(2005) 010603.

\bibitem{Wol10}
U.\ Wolff, Nucl.\ Phys.\ B832 (2010) 520.

\bibitem{Lue82a}
M.\ L\"uscher, Nucl.\ Phys.\ B205 (1982) 483.

\bibitem{Phi86}
A.\ Phillips and D.\ Stone, Commun.\ Math.\ Phys.\ 103 (1985) 599.

\bibitem{Goe86}
M.\ G\"ockeler, M.\ L.\ Laursen, G.\ Schierholz, and U.-J.\ Wiese, 
Commun.\ Math.\ Phys.\ 107 (1986) 467.

\bibitem{Goe87}
M.\ G\"ockeler, A.\ S.\ Kronfeld, M.\ L.\ Laursen, G.\ Schierholz, and
U.-J.\ Wiese, Nucl.\ Phys.\ B292 (1987) 349.

\bibitem{Pug89}
D.\ Pugh and M.\ Teper, Phys.\ Lett.\ B324 (1989) 159.

\bibitem{Goe89}
M.\ G\"ockeler, A.\ S.\ Kronfeld, M.\ L.\ Laursen, G.\ Schierholz, and
U.-J.\ Wiese, Phys.\ Lett.\ B233 (1989) 192.

\end{thebibliography}
\end{document}